\documentclass[
reprint,
superscriptaddress,
amsmath,amssymb,
aps,
prl,
]{revtex4-2}
\usepackage{CJK}
\usepackage{array}
\usepackage{graphicx}
\usepackage{dcolumn}
\usepackage{bm}
\usepackage{xcolor}
\usepackage{colortbl}
\usepackage{ulem} 
\usepackage{soul}
\usepackage{mathrsfs}
\usepackage{lipsum}
\usepackage{hyperref} 
\usepackage{cleveref} 
\usepackage{marginnote}
\hypersetup{colorlinks = true, 
	    linkcolor = blue, 
	    urlcolor = blue,
            citecolor = blue} 

\usepackage{pifont}
\usepackage{ulem}
\usepackage{cancel}

\begin{document}
\begin{CJK*}{UTF8}{gbsn}
\preprint{First Draft}
\title{Topological phase transition in chaotic optomechanical systems}
\author{Xiao-Jun Zhang ({\CJKfamily{gbsn}张晓军})}
\email{zhangxj037@nenu.edu.cn}
\affiliation{School of Physics and Center for Quantum Sciences, Northeast Normal University, Changchun 130024, China}

\begin{abstract}
Hidden structures with well-defined predictability are uncovered in the evolution of a chaotic optomechanical system from the perspective of the $\epsilon$-machine. Tuning the frequency of the driving laser can switch off this predictability, and such behaviour corresponds to a phase transition that is deeply related to topological changes in phase space. The transition probabilities between causal states allow us to define an entropy (uncertainty) that serves as an effective order parameter. This phase transition can be readily demonstrated in currently available experiments by monitoring the quadrature of the optical mode. We hope that this work could fundamentally broaden the regimes of cavity micromechanics and nonlinear optics.
\end{abstract}
\maketitle
\end{CJK*}

\section{Introduction}
\begin{figure}[b]
\centering
\includegraphics[width= 8 cm]{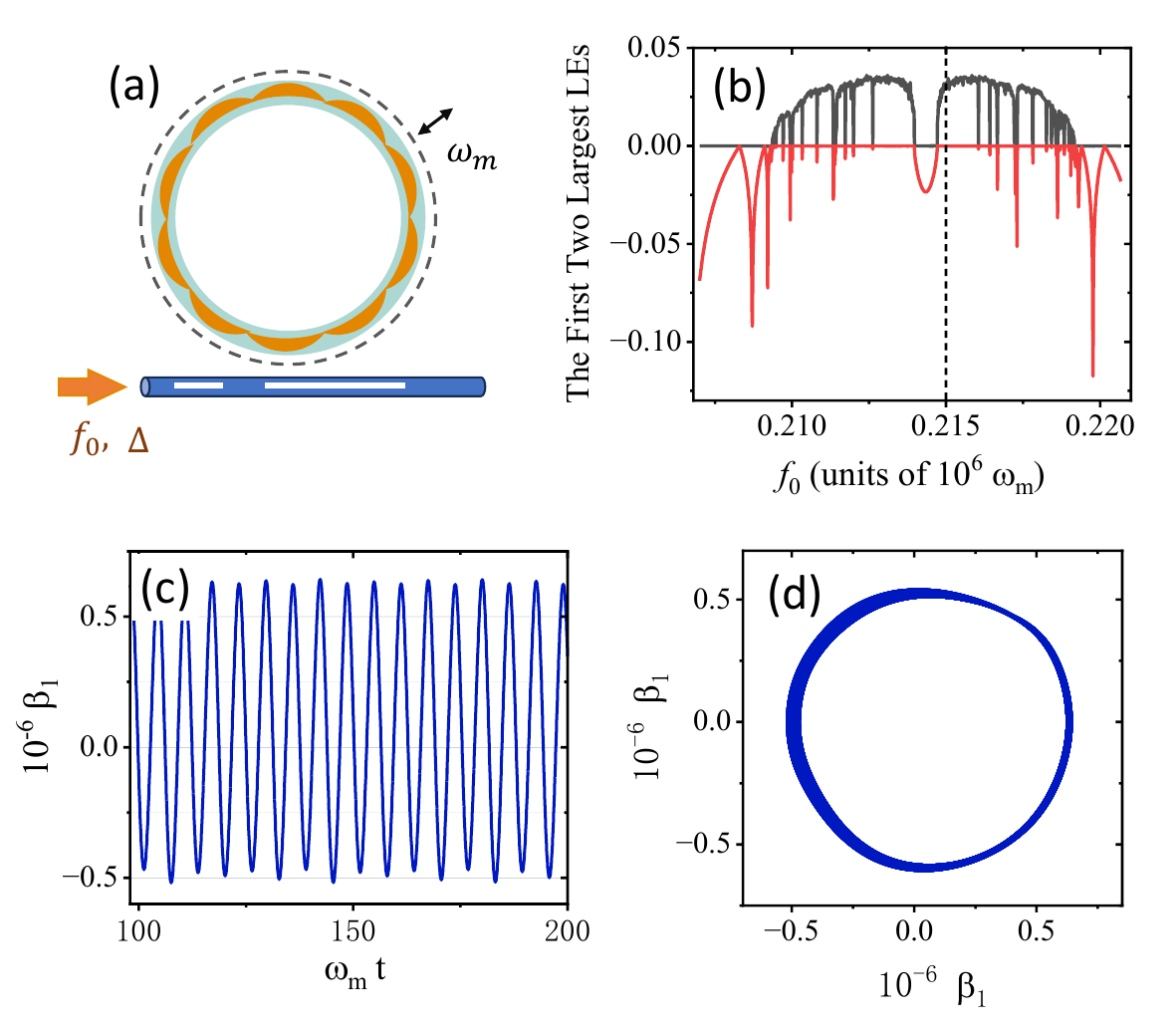}
\caption{\label{figA} (a) Schematic diagram of an optomechanical system consist of optical resonator with mechanical breathing mode and a tapered fiber. (b) The largest and the second largest Lyapunov exponents under different pumping strength $f_0$ with $f_0 = 0.2150\times 10^6\,\omega_m$ marked by the vertical dashed line. (c) The position of the mechanical oscillator $\beta_1$ evolving with respect to time and $f_0 = 0.2150\times 10^6\,\omega_m$. (d) Trajectory projected onto the phase space of the mechanical mode with $f_0 = 0.2150\times 10^6\,\omega_m$. Other parameters are set as 
$\Delta = 1.00 \,\omega_m$, 
mass and radius of the resonator are $m = 10.0$ ng and $R = 1.1$ mm. $\lambda_c = 1.55\,\mu$m.  
$G = 7.1712 \times 10^{-6} \omega_m$, 
$\kappa = \omega_c / (3.2\times 10^7) \simeq 0.6026 \,\omega_m$, 
$\gamma_m = 8.2752\times10^{-5}\,\omega_m$ 
and $\omega_m = 6.3020 \times 10^7$ Hz. }
\end{figure}

An optomechanical system, where light couples with mechanical motion, are typically realized by a laser interacting with a movable mirror or vibrating microresonator \cite{RevModPhys.86.1391}, see Fig. \ref{figA}(a). Radiation pressure or photothermal forces let light drive \cite{PhysRevLett.95.033901}, cool \cite{WOS:000292911200038,WOS:000295575400040} the mechanical motion, enabling precision sensing \cite{PhysRevLett.130.093603} and quantum control \cite{WOS:000448900900044}.
Intracavity photon-mechanical interaction yields nonlinear coupling, giving rise to chaos \cite{PhysRevLett.114.013601} and hyperchaos  \cite{rv1f-x73d}.
The advantage of the optomechanical system, such as tunable interaction,  high-Q resonators and sensitive readout, allows people to extend the research of the optical chaos into various regimes \cite{PhysRevLett.114.253601,PhysRevA.101.053851,PhysRevA.104.033522,PhysRevA.104.L031503,PhysRevA.104.023525,PhysRevA.107.033522}.

Chaos is deterministic but shows sensitive dependence on initial conditions. Tiny changes in initial state grow exponentially, making the long-term trajectories practically unpredictable \cite{Lorenz1963}.
In this letter, we theoretically demonstrate that, under specific parameter conditions, chaos in optomechanical systems possesses inherent structures concealed beneath the apparent randomness. It can be regarded as predictable within the framework of computational or statistical mechanics. The shift from a predictable to an unpredictable evolution belongs to the category of phase transitions.

The concepts we borrow from computational mechanics are the causal state and
$\epsilon$-machine \cite{WOS:000170860400011}. After coarse-graining the mechanical position into a time series, we partition it into past histories and corresponding futures. Grouping together histories that induce the same conditional distribution over futures defines equivalence classes. The minimal predictive macrostates (causal states) and their transition probabilities define the $\epsilon$-machine  \cite{WOS:000245960700016, mycomment1}, and it reveal the hidden structures. The transition probabilities allow us to define a order parameter to distinguish phases, and the phase transition related intrinsically to the topological changes of the phase-space geometry.

In general, the Hamiltonian of an optomechanical system reads
$H/\hbar = \Delta a^\dagger a + \omega_m b^\dagger b - G a^\dagger a (\hat{b} + \hat{b}^\dagger) + i f_0(\hat{a}^\dagger - \hat{a})$, where $\hat{a}$ ($\hat{b}$) is the annihilation operator of the optical (mechanical) mode, and $\omega_m$ is the resonant frequency of the oscillator. The strength of the optical-mechanical interaction is $G$. A laser is driving the cavity at frequency $\omega_L$ with the strength represented by $f_0$, and it can be translate into the power of the laser by $P = (f_0^2 / 2 \kappa) \hbar \omega_L$. $\Delta$ is the detuning defined as $\Delta = \omega_c - \omega_L$ where $\omega_c$ is the resonant frequency of the optical cavity.
Under the semiclassical approximation where the photon-phonon quantum correlation is neglected, we can set $\langle\hat{a}\rangle = \alpha$ and $\langle \hat{b}\rangle = \beta$. The real and imaginary parts of $\alpha$ and $\beta$ from $\alpha = \alpha_1 + i \alpha_2$ and $\beta = \beta_1 + i \beta_2$ evolve with respect to time as
\begin{subequations} \label{dyeq}
\begin{align}
\frac{d \alpha_1}{dt} &= -\kappa \alpha_1 + (\Delta - 2 G \beta_1) \alpha_2 + f_0 \label{dyeq1}\\
\frac{d \alpha_2}{dt} &= -\kappa \alpha_2 - (\Delta - 2 G \beta_1) \alpha_1 \label{dyeq2}\\
\frac{d \beta_1}{dt} &= - \gamma_m \beta_1 + \omega_m \beta_2\label{dyeq3}\\
\frac{d \beta_2}{dt} &= - \gamma_m \beta_2 - \omega_m \beta_1 + G(\alpha_1^2 +\alpha_2^2)\label{dyeq4}
\end{align}
\end{subequations}
The value of $\beta_1$ represents the position of the mechanical oscillator. With carefully arranged pumping, for example, $\Delta = 1.00\, \omega_m$ and around $f_0 = 0.215\times10^6\, \omega_m$, the system become chaotic. As shown in Fig. \ref{figA}(b) where the first and second largest Lyapunov exponents are plotted. The largest Lyapunov exponent larger than 0 indicates the system become chaotic. Since our system is a smooth autonomous continuous-time system, at least one Lyapunov exponent is guaranteed to be zero. The position $\beta_1$ oscillates at a frequency approximate to $\omega_m$, as shown in Fig. \ref{figA}(c). However, the local maximums/peaks never rest on the same value, so does the minimum (maybe more obvious compared with the peaks).
This leads to a shape of strip circle in the phase space see Fig. \ref{figA}(d), indicating that the long-time accurate prediction is difficult.

\begin{figure}[t]
\centering
\includegraphics[width= 8.5 cm]{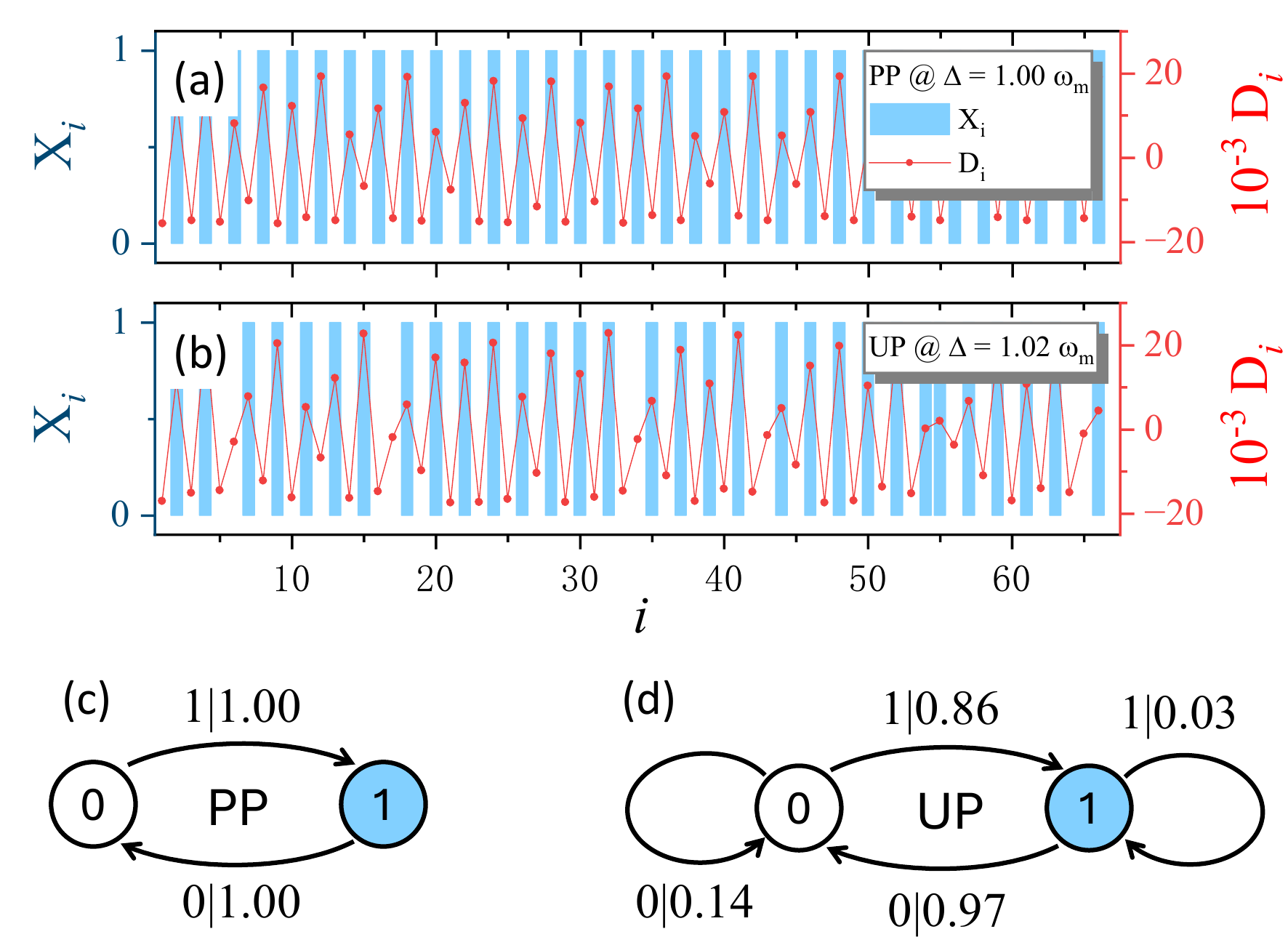}
\caption{\label{figB} 
Discretization of $\beta_1$ realized by calculating the difference $D_i$ between the adjacent local maxima (red broken line) and its binarization: $X_i = 1$ if $D_i \geq 0$ (bars), $X_i = 0$ if $D_i < 0$ (empty), in PP with $\Delta = 1.00\,\omega_m$ 
(a) and in UP with $\Delta = 1.02\,\omega_m$ (b). 
The corresponding $\epsilon$-machines are presented respectively as (c) and (d). 
Transitions between the causal states are represented by arrow with the relevant transition probabilities. For example, arrow with label ``$0|0.97$'' means $T_{10}^{(0)} = 0.97$. The driving coefficient is set as
$f_0 = 0.215\times 10^6\,\omega_m$, and other parameters are identical with that in Fig.~\ref{figA}.}
\end{figure}

\begin{figure}[t]
\centering
\includegraphics[width= 8.5 cm]{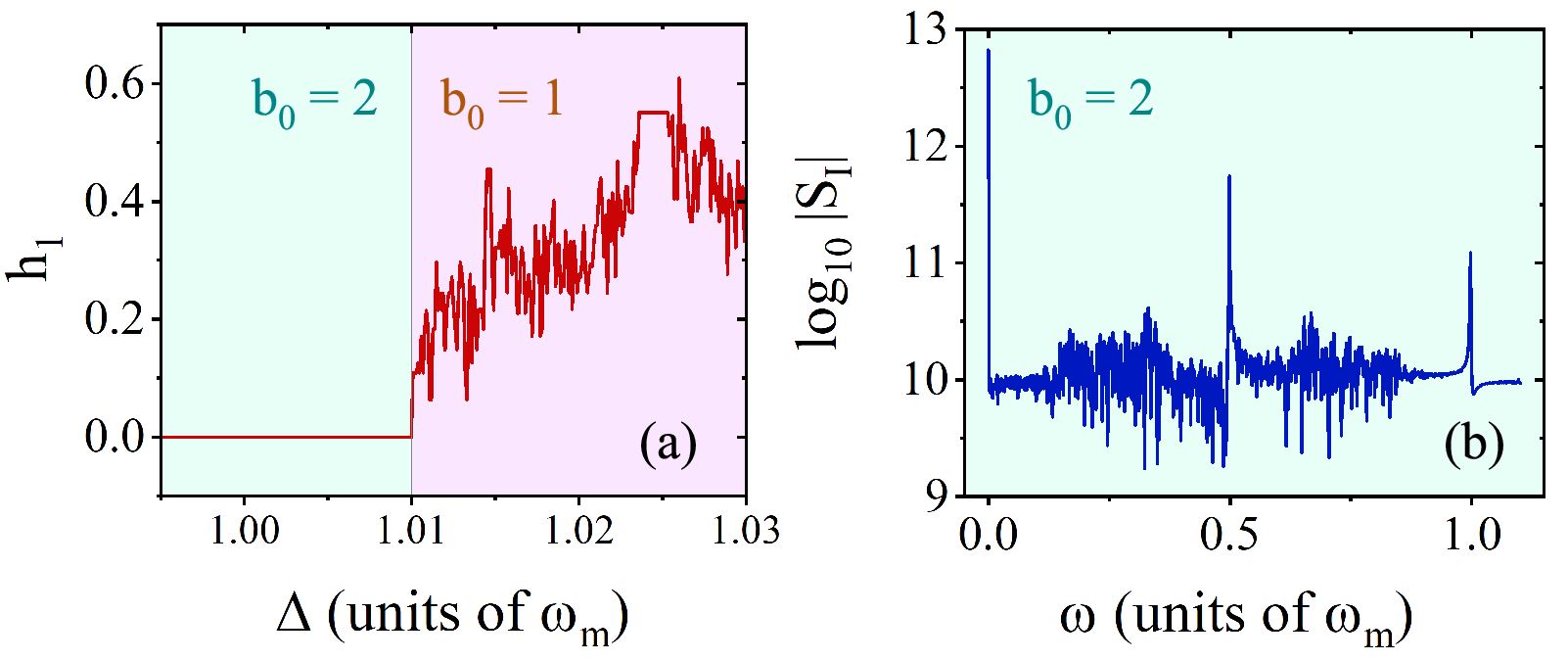}
\caption{\label{figC}
(a) The uncertainty $h_1$ plotted against the detuning $\Delta$, the light-green backgound marks the regime with first Betti number $b_1 = 2$, and light-pink background corresponds to $b_1 = 1$. (b) The frequency spectra $\log_{10}|S_I(\omega)|$ in PP with $\Delta = 1.00 \,\omega_m$. The peaks at $0.5\,\omega_m$ and $\omega_m$ are responsible for the hidden structure and the fundamental oscillation of the mechanical mode.
We set $f_0 = 0.215\times 10^6\,\omega_m$, and other parameters are identical with that in Fig. \ref{figA}.}
\end{figure}

\begin{figure*}[t]
\centering
\includegraphics[width= 17.5 cm]{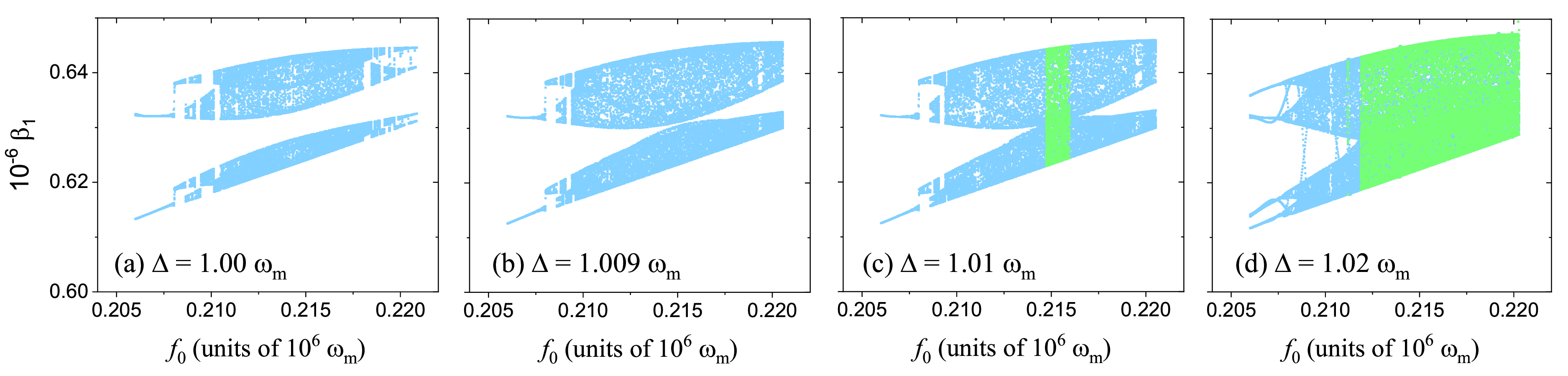}
\caption{\label{figD} 
Bifurcation diagram of $\beta_1$ with respect to the driving coefficient $f_0$ for different detuning $\Delta$: (a) $\Delta = 1.00 \,\omega_m$, (b) $\Delta = 1.009 \,\omega_m$, (c) $\Delta = 1.01 \,\omega_m$ and (d) $\Delta = 1.03 \,\omega_m$
$\kappa = 0.6\,\omega_m$, the blue colored regin corresponds to the cases with $b_0 = 2$, while the green color is for $b_0$ = 1.}
\end{figure*}

To make an effective prediction over the evolution of $\beta_1$, discretization are applied by only taking account of, e.g., the maximal value of $\beta_1$, and one gets a series of peaks $A_i$. It is then converted into a binary series $X_i$ by the
following process: Define $D_i = A_{i+1} - A_{i}$, then $X_i = \theta(D_i)$, where $\theta$(x) is a Heaviside step function. $X_n$ gets the value of 1 if the next peak is increased and does the value of 0 otherwise. We can treat $X_n$ as a language spoken by the optomechanical system, and its alphabet is consist of $0$ and $1$. 

As shown in Fig. \ref{figB}(a) with $\Delta = 1.00 \, \omega_m$, the pattern is obvious.  The peaks of $\beta_1$ rise and fall alternately. Nevertheless, the exact value of \(\beta_1\) remains stochastic. Consequently, precise quantitative predictions cannot be achieved, but the evolving trend can be reliably identified. From this perspective,  we call it that the system is in a predictable phase (PP). 

Under the framework of $\epsilon$-machine, the behaviour of $\beta_1$ can be well predicted by two causal states $s_0$ and $s_1$. The state $s_0$ is defined by a one-bit effective history $\overleftarrow{x}=\cdots 0$, while $s_1$ is associated with  $\overleftarrow{x}=\cdots 1$.
The transition probability from $s_0$ to $s_1$ while emitting symbol $1$ is $T_{0,1}^{(1)}$, and it always takes the value of $1$, similarly $T_{1,0}^{(0)} = 1$. They are represented by the arrows in Fig. \ref{figB}(c). The other two transition probabilities $T_{0,1}^{(0)}$ and $T_{1,0}^{(1)}$ are zeros. In other words, the sequence of $\cdots 00$ and $\cdots 11$ never appears.

The situation is changed when $\Delta = 1.02\,\omega_m$. Such determinism vanishes, as we can see that in Fig. \ref{figB}(b), the position $\beta_1$ could be increased/decreased at two adjacent maximums. Consequently, two additional nonzero transitions appear in the $\epsilon$-machine, see Fig. \ref{figB}(d). We call it that the system is in an unpredictable phase (UP) \cite{mycomment2}.

Without the determinism, UP exhibits time translational symmetry: shifting it by an arbitrary time period yields a $X_i$ sequence identical to the original unshifted sequence in the sense of randomness. Reducing $\Delta$ from $1.02$ to $1.00\,\omega_m$, the optomechanical system undergoes spontaneous symmetry breaking, and structures (periods) in time dimension appears.
The behaviour of the system resembles that of time crystals \cite{WOS:000837874700046}, as the system responds to a driving signal of period $T$ with a periodic output of period $2T$. The distinction is that this response is covered in chaotic oscillations.
The formation of the structure is accompanied by the reduction of local randomness. We can quantify such changes using the probability of finding the state $s_i$, namely $P(s_i)$, and the transition probability $T_{ij}^{a}$ by defining the uncertainty as
\[
h_1 = -\sum_{i} \sum_{j} 
P(s_i) T_{ij}^{(j)} \log_2 T_{ij}^{(j)}.
\]
For the case of PP, $h_1 = 0$, meaning that we can be absolutely sure about the future. By tuning $\Delta$, as shown in Fig. \ref{figC}(a) that, once it exceeds a certain threshold, this quantity jumps to a nonzero value, indicating that the system enters the an UP state. For the parameters used in Fig. \ref{figC}(a), this critical point is around $\Delta = 1.01\,\omega_m$. Since we calculate $h_1$ over finite time period, it only reflects the short term randomness and the numerical values fluctuate.

The transition from the unpredictable phase to the predictable phase can be illustrated more vividly by examining the bifurcation diagram. For $\Delta = 1.00 \, \omega_m$, PP state exhibits a two-band structure, see Fig. \ref{figD}(a). This results from the dynamics of the mechanical mode. Driven by the cavity field, the mechanical mode evolves with respect to time according to a damped driven oscillator. Eqs. (\ref{dyeq3}) and (\ref{dyeq4}) suggest that
\begin{equation}
\begin{split}
&\frac{d^2}{dt^2}\beta_1 + 2 \gamma_{m} \frac{d}{dt}\beta_1 + (\omega_m^2+\gamma_m^2) \beta_1 =  \omega_m G I(t),
\end{split}
\end{equation}
with $I(t) = \alpha^*\alpha$ being the intensity of the classical cavity mode.
Based on Eqs. (\ref{dyeq1}) and(\ref{dyeq2}), the cavity mode $\alpha$ oscillates at frequency of $\Omega = \Delta-2G\beta_1(t)$. We can use a sinusoidal ansatz to represent $\beta_1(t)$, explicitly, $\beta_1(t) = B(t) \cos (\omega t)$, where $B(t)$ is the time-varying amplitude and $\omega \simeq \omega_m$. Then, the intensity
$I(t)$ can be expanded as $I(t) \simeq I_0 + I_1 \cos(\frac{\omega_m}{2} t) + I_2 \cos (\omega_m t) +\cdots$. The strong component at $\frac{\omega_m}{2}$ from the driving force, leads to a subharmonic (1/2) response. In the time domain, this subharmonic response yields a waveform that repeats every two cycles of the fundamental oscillation with alternating peak amplitudes. Figure~\ref{figC}(b) displays $S_I(\omega)$, the Fourier spectrum of intensity $I(t)$, revealing spectral components close to $\omega_m/2$ and $\omega_m$. 

The essential of the statistic certainty is the transition between the two band. The state of the system ``travels'' in the four-dimensional phase space spanned by $\alpha_1$, $\alpha_2$, $\beta_1$ and $\beta_2$, but always visits the two bands in the $\beta_1$-$\beta_2$ subspace in the same order.
The growing detuning $\Delta$ changes the shapes of the band, and around the critical point, the two band touches, see Fig. \ref{figD}(b) and Fig. \ref{figD}(c), and futher large $\Delta$ lead to a band-merging crisis \cite{GREBOGI1983181,Ott_2002}, after coalescence of two chaotic bands into a single connected chaotic attractor, the hidden structure disappears.

Transition between PP and UP states is intrinsically associated with topological variations in phase space. 
For different driving power, the two band merge at different $\Delta$. At this point, zeroth Betti number $b_0$ which is the rank of the $0^{th}$ singular homology group $H_0(X)$ can be a workable indicator. If a set is connected, $b_0 = 1$, and if it has two disconnected pieces, $b_0=2$, and so on. We numerically obtained the value of $b_0$, and find the transition from $b_0 = 2$ (green background in Fig. \ref{figC}(a)) to $b_0 = 1$ (red backgound) at around $\Delta = 1.01 \,\omega_m$, identical to the threshold where abrupt change takes place in uncertainty. 
Thus, both the uncertainty and the zeroth Betti number qualify as suitable order parameters for characterizing the corresponding topological phase transition.

\begin{figure}[t]
\centering
\includegraphics[width= 8 cm]{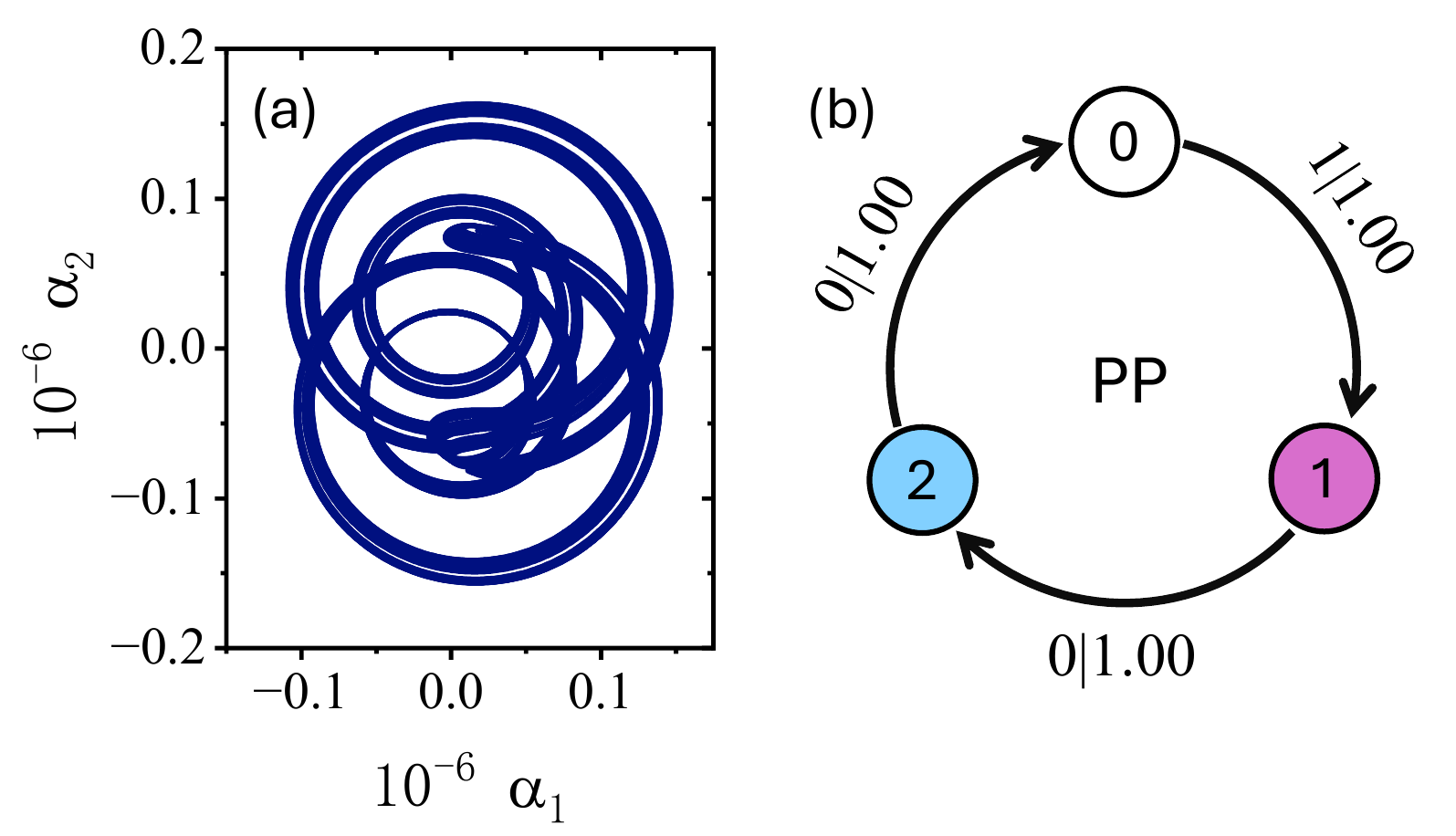}
\caption{\label{figE} 
Trajectory projected on subspace $\alpha_1$-$\alpha_2$ and the corresponding $\epsilon$-machine built from the motion of $\alpha_1$. Casual state $s_0$ is define by history 
$\overleftarrow{x}=\cdots 00$, $s_1$ defined by $\overleftarrow{x}=\cdots 01$ and $s_2$ defined by $\overleftarrow{x}=\cdots 10$. We set $f_0 = 0.215\times 10^6\,\omega_m$, $\Delta = 1.00 \,\omega_m$ and other parameters are identical with that in Fig.~\ref{figA}.}
\end{figure}

Since the phase transition does not depends on other auxiliary system, the author speculates that it might not be difficult to observe it in the microtoroid/microring whispering-gallery resonator with mechanical radial breathing mode. Possible experiments platforms could be the one reported in letter  \cite{PhysRevLett.98.167203} where chaotic behaviour is experimentally demonstrated. The parameters we select are close to those used in the experiments reported in Ref. \cite{PhysRevLett.104.083901,WOS:000436594300056}; thus their experimental platforms are promising candidates as well.  In general, the mechanical motion of $\beta_1(t)$ can be inferred from measuring the output-field quadrature which corresponds to the cavity amplitude quadrature $\alpha_1(t)$, using homodyne/heterodyne detection. The motion of light mode could be complex than the mechanical one. For the parameters we choose, the trajectory in phase space $\alpha_1$-$\alpha_2$ shows sophisticated patten (Fig. \ref{figE}(a)). However, determinism still persists. Following the exactly same procedure of discretization and analysis, we find that $\alpha_1$ at $\Delta = 1.00 \,\omega_m$ can be well predicted with a three-causal-state $\epsilon$-machine, as shown by Fig. \ref{figE}(b), with transition probability being either 1 or 0. Such feature leads to $h_1 = 0$, just as that in mechanical mode.

We acknowledge financial support from the industrial collaborative
project (Project No. 103-401125103, Contract No. 2025-
2200-0200-0084).

\bibliography{refs.bib}

\begin{thebibliography}{25}
\bibitem [{myc({\natexlab{a}})}]{mycomment1}%
  \BibitemOpen
  \href@noop {} {} ({\natexlab{a}}),\ \bibinfo {note} {a breaf introduction to
  causal state, transition probability and $\epsilon$-mechine can be found in
  Appendix}\BibitemShut {NoStop}%
\bibitem [{myc({\natexlab{b}})}]{mycomment2}%
  \BibitemOpen
  \href@noop {} {} ({\natexlab{b}}),\ \bibinfo {note} {we employ the history
  length of the PP state to construct an \(\epsilon\)-machine corresponding to
  the UP state, so as to characterize abrupt transitions from the PP state to
  the UP state. For the definition of the proper history length, see
  Appendix}\BibitemShut {NoStop}%
\end{thebibliography}

\section{Appendix: Brief introduction to $\epsilon$-machine}


For a bi-infinite sequence sequence of random variables $\{\cdots, X_{-1}, X_{0}, X_{1}, \cdots\}$, each $X_t$ takes a value from a finite alphabet of size $k$. At any given time $t$, the semi-infinite part $\overleftarrow{X}_t = \{\cdots, X_{t-1}, X_{t}\}$ is call a history, while the other half $\overrightarrow{X}_t = \{ X_{t+1}, X_{t+2},\cdots\}$ is its corresponding future. Two realizations of the history $\overleftarrow{x}$ and $\overleftarrow{x}'$ are considered equivalent, if and only if their conditional probability distribution for the entire future are identical 
\[\mathrm{Pr}(\overrightarrow{X}|\overleftarrow{X} = \overleftarrow{x}) = \mathrm{Pr}(\overrightarrow{X}|\overleftarrow{X} = \overleftarrow{x}').\] A causal state $s_i$ is a set of all histories that lead to the same probabilities prediction of the future. It is the smallest sufficient statistic for predicting the future of the process.

The Shannon entropy of the stationary distribution over causal states is call statistical complexity $C_\mu$, 
\[C_\mu = -\sum_i \mathrm{Pr}(s_i) \log_2 \mathrm{Pr(s_i)},\] where $\mathrm{Pr}(s_i)$ means the probability of causal state $s_i$. It is the average number of bits of memory the $\epsilon$-machine must store about the current causal states to predict optimally.

\begin{figure}[b]
\centering
\includegraphics[width= 8 cm]{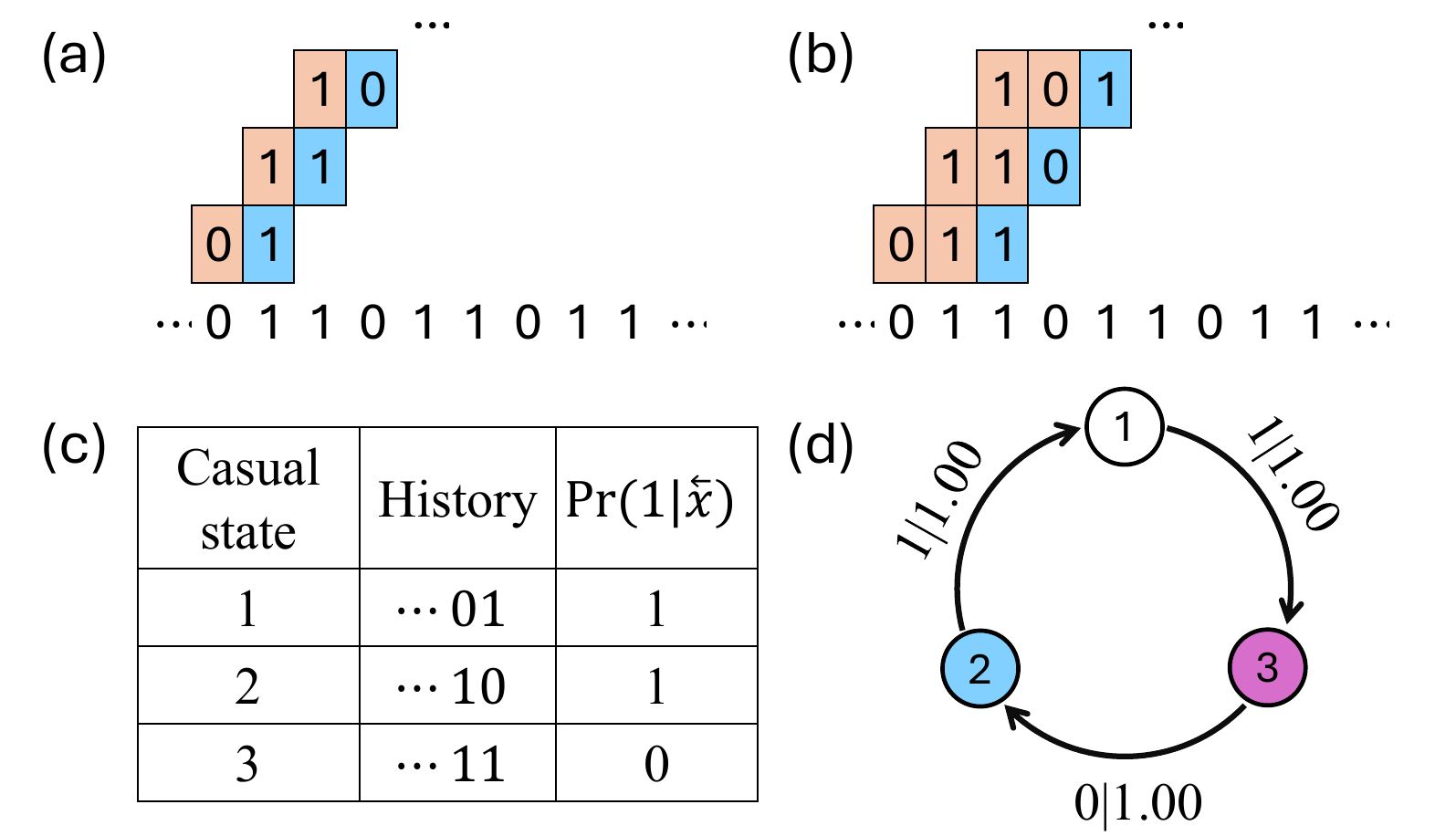}
\caption{\label{figF} 
Trajectory projected on subspace $\alpha_1$-$\alpha_2$ and the corresponding $\epsilon$-machine built from the motion of $\alpha_1$. Casual state $s_0$ is define by history 
$\overleftarrow{x}=\cdots 00$, $s_1$ defined by $\overleftarrow{x}=\cdots 01$ and $s_2$ defined by $\overleftarrow{x}=\cdots 10$. We set $f_0 = 0.215\times 10^6\,\omega_m$, $\Delta = 1.00 \,\omega_m$ and other parameters are identical with that in Fig.~\ref{figA}.}
\end{figure}

A simple example is a process that repeatedly generating ``011'', that is $\cdots 011011011 \cdots$. For simplify, we only focus on the predication made by only examining a one-bit future. The numerical algorithm for finding the statistical complexity, normally starts the analysis by first taking a one-bit history, as shown in Fig. \ref{figF}(a). Shifting the window one bit at a time, to get the corresponding history and future, then calculate the future distribution for each history, collecting and classifying them into several causal states to find the value of $C_\mu$. Then try the two-bit history, see Fig. \ref{figF}(b), to obtain a new value of $C_\mu$, then deal with the three-bit history, and so on. Until the statistical complexity converges to certain value, the analysis stops, and the proper length of the history is found. Practically, the length of history to be considered is limited by $L_{max} < \mathrm{log}_k N$ where $N$ is the total number of the data point. In our example, $k = 2$.

For our example, It is easy to find out that the two-bit history is good enough, and three histories are found, they are
$\overleftarrow{x}_1 = \cdots 01$, $\overleftarrow{x}_2 = \cdots 10$ and $\overleftarrow{x}_3 = \cdots 11$.
The conditional probability $\mathrm{Pr}(1|x_1)$, that is the probability for ``1'' generated following $\overleftarrow{x}_1$, equals 1. Additionally, $P(1|x_2) = 1$ and $P(1|x_3) = 0$.

The transition probability $T_{ij}^{(a)}$ is defined as the probability of emitting a symbol \(a\) while transitioning from state \(s_i\) to state \(s_j\), i.e.,
\[
T_{ij}^{(a)} = \Pr (\overleftarrow{X}a \in s_j \mid \overleftarrow{X}\in s_i).
\]
By simple observation, one find that, in our example, histories $\overleftarrow{x}_1$ and $\overleftarrow{x}_2$ have the same the same distribution of conditional probabilities for all possible future events. However we cannot categorize them as the same causal state.
The reason is that, in an $\epsilon$-machine, transition between causal states should be deterministic: for each causal state and emitted symbol there is a unique next causal state. If merging $x_1$ and $x_2$ would violate this property. As we can see that $x_1$ followed by 1 becomes $s_3$, and $x_2$ followed by 1 becomes $s_1$. Based on this fact, we treat $\overleftarrow{x}_1$ and $\overleftarrow{x}_2$ as histories of two different casual states, as shown in Fig. \ref{figF}(c). And the corresponding $\epsilon$-machine is plotted as Fig. \ref{figF}(d).

\end{document}